\documentclass[12pt,preprint]{emulateapj}

\RequirePackage{color}

\newcommand{\gsim}{\hspace{0.3em}\raisebox{0.4ex}{$>$}\hspace{-0.75em}\raisebox{-.7ex}{$\sim$}\hspace{0.3em}}

\shortauthors{Homma et al.}
\shorttitle{A new satellite in the Milky Way}

\begin{document}

\title{A New Milky Way Satellite Discovered In The Subaru/Hyper Suprime-Cam Survey}

\author{Daisuke~Homma\altaffilmark{1}, Masashi~Chiba\altaffilmark{1}, 
Sakurako~Okamoto\altaffilmark{2}, Yutaka~Komiyama\altaffilmark{3,4}, 
Masayuki~Tanaka\altaffilmark{3}, Mikito~Tanaka\altaffilmark{1},
Miho~N.~Ishigaki\altaffilmark{5}, Masayuki~Akiyama\altaffilmark{1},
Nobuo~Arimoto\altaffilmark{6,4},
Jos\'e~A.~Garmilla\altaffilmark{7}, Robert~H.~Lupton\altaffilmark{7},
Michael~A.~Strauss\altaffilmark{7},
Hisanori~Furusawa\altaffilmark{3},
Satoshi~Miyazaki\altaffilmark{3,4},
Hitoshi~Murayama\altaffilmark{5},
Atsushi~J.~Nishizawa\altaffilmark{8},
Masahiro~Takada\altaffilmark{5},
Tomonori~Usuda\altaffilmark{3,4}, and
Shiang-Yu~Wang\altaffilmark{9}
}

\altaffiltext{1}{Astronomical Institute, Tohoku University, Aoba-ku,
Sendai 980-8578, Japan \\E-mail: {\it d.homma@astr.tohoku.ac.jp}}
\altaffiltext{2}{Shanghai Astronomical Observatory, 80 Nandan Road, Shanghai 200030, China}
\altaffiltext{3}{National Astronomical Observatory of Japan, 2-21-1 Osawa, Mitaka,
Tokyo 181-8588, Japan}
\altaffiltext{4}{The Graduate University for Advanced Studies, Osawa 2-21-1, Mitaka, Tokyo 181-8588, Japan}
\altaffiltext{5}{Kavli Institute for the Physics and Mathematics of the Universe (WPI),
The University of Tokyo, Kashiwa, Chiba 277-8583, Japan}
\altaffiltext{6}{Subaru Telescope, National Astronomical Observatory of Japan, 650 North A'ohoku Place,
Hilo, HI 96720, USA}
\altaffiltext{7}{Princeton University Observatory, Peyton Hall, Princeton, NJ 08544, USA}
\altaffiltext{8}{Institute for Advanced Research, Nagoya University, Furo-cho,
Chikusa-ku, Nagoya 464-8602, Japan}
\altaffiltext{9}{Institute of Astronomy and Astrophysics, Academia Sinica, Taipei, 10617, Taiwan}

\begin{abstract}
We report the discovery of a new ultra-faint dwarf satellite companion of the Milky Way 
based on the early survey data from the Hyper Suprime-Cam Subaru Strategic Program.
This new satellite, Virgo~I, which is located in the constellation of Virgo,
has been identified as a statistically significant (5.5$\sigma$) spatial overdensity
of star-like objects with a well-defined main sequence and red giant branch
in their color-magnitude diagram. The significance of this overdensity
increases to 10.8$\sigma$ when the relevant isochrone filter is adopted for the search.
Based on the distribution of the stars around the likely main sequence turn-off at
$r \sim 24$ mag,
the distance to Virgo~I is estimated as 87~kpc, and its most likely
absolute magnitude calculated from a Monte Carlo analysis is
$M_V = -0.8 \pm 0.9$~mag. This stellar system has an extended
spatial distribution with a half-light radius of 38$^{+12}_{-11}$~pc, which clearly
distinguishes it from a globular cluster with comparable luminosity. Thus, Virgo~I
is one of the faintest dwarf satellites known and is located beyond the reach
of the Sloan Digital Sky Survey. This demonstrates the power of this
survey program to identify very faint dwarf satellites. This discovery of Virgo~I
is based only on about 100 square degrees of data, thus a large number of
faint dwarf satellites are likely to exist in the outer halo of the Milky Way.
\end{abstract}


\section{Introduction}

Dwarf spheroidal galaxies (dSphs) associated with the Milky Way (MW) and Andromeda
galaxies provide important constraints on the role of dark matter in galaxy
formation and evolution. Indeed, these faint stellar systems are largely dominated
by dark matter with mass-to-luminosity ratios of 10 to 1000 or even larger in
fainter systems, based on their stellar dynamics \citep{Gilmore2007,Simon2007}.
Thus, the basic properties of dSphs, such as their total number and
spatial distributions inside a host halo like the MW, provide
useful constraints on dark matter on small scales, in particular the nature
and evolution of cold dark matter (CDM) in a $\Lambda$ dominated universe.

One of the tensions between theory and observation is the missing satellite
problem: the theory predicts a much larger number of subhalos in a MW-like halo
than the observed number of satellite galaxies \citep{Klypin1999,Moore1999}.
Solutions to this problem are to consider other types of dark matter
than CDM \citep[e.g.,][]{Maccio2010}
or to invoke baryonic physics \citep[e.g.,][]{Sawala2016}.
Another possibility is that we have seen only a fraction of all the
satellites associated with the MW due to various observational biases \citep{Tollerud2008}. 
Motivated by this, a systematic search for new dSphs has been made based on
large survey programs, such as the Sloan Digital Sky Survey (SDSS) \citep{York2000}
and the Dark Energy Survey (DES) \citep{Abbott2016}. SDSS discovered
15 ultra-faint dwarf galaxies (UFDs) with $M_V \gsim -8$ mag
\citep[e.g.,][]{Willman2005,Sakamoto2006,Belokurov2006},
and DES recently reported the discovery of many more candidate UFDs
in the south \citep[e.g.,][]{Bechtol2015,Koposov2015,Drlica-Wagner2015}.
These discoveries are consistent with the work by \citet{Tollerud2008},
anticipating that there exists a large number of yet unidentified dwarf satellites
in the MW halo, especially in its outer parts.

This paper reports the discovery of a new faint dwarf satellite in the MW,
in the course of the Subaru Strategic Program (SSP) using Hyper Suprime-Cam (HSC).
HSC is a new prime-focus camera on the Subaru telescope with a 1.5~deg diameter
field of view \citep{Miyazaki2012}, which thus allows
us to survey a large volume of the MW halo out to
a large distance from the Sun, where
a systematic search for new satellites has not yet been undertaken.

\begin{figure*}[t!]
\centering
\includegraphics[width=120mm]{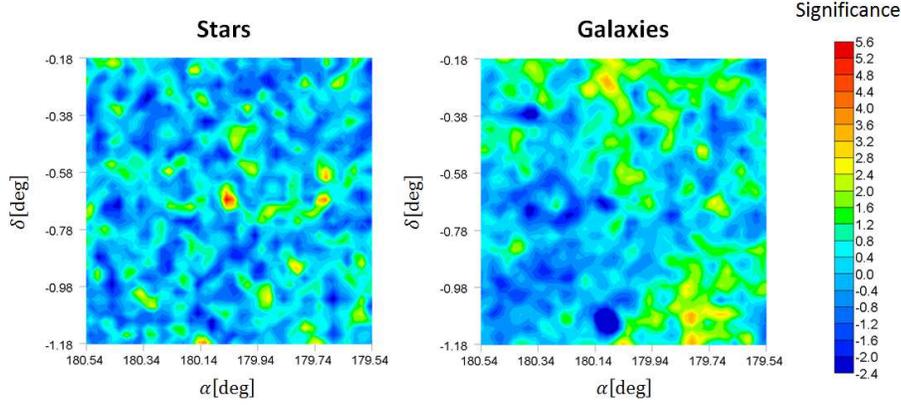}
\caption{
Left panel: the spatial distribution of the sources classified as stars with $i < 24.5$ mag
and $g-r<1.0$, covering one square degree centered on the candidate
overdensity of stars. The star counts are in bins of $0^{\circ}.05 \times 0^{\circ}.05$.
Right panel: the plot for the sources classified as galaxies with $i < 24.5$ mag
and $g-r<1.0$. Note that there is no overdensity at the center of this plot.
}
\label{fig: space}
\end{figure*}

\begin{figure*}[t!]
\centering
\includegraphics[width=150mm]{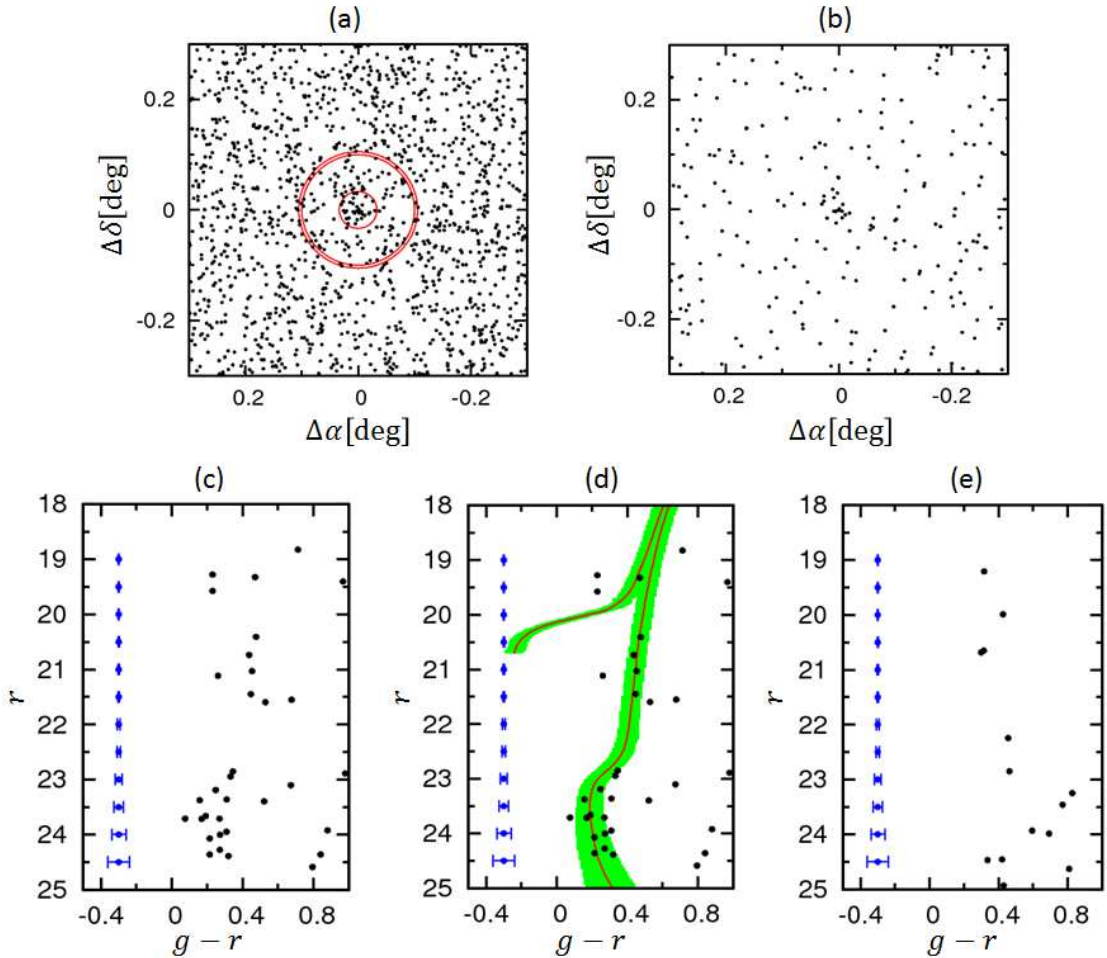}
\caption{
The spatial distribution of the stars around the overdensity (upper panels, where $\Delta\alpha$
and $\Delta\delta$ are the relative offsets in celestial coordinates) and their distribution in
the $g-r$ vs. $r$ CMD (lower panels).
Panel (a): spatial distribution of the sources classified as stars with $i < 24.5$ mag
and $g-r<1.0$.
Red circles denote annuli with radii $=2'$, $6'$, and $6'.33$ from the center.
There is an overdensity around the field center with statistical significance of 5.5$\sigma$.
Panel (b): the same as (a) but for the stars passing the isochrone filter shown in panel (d).
The statistical significance of the overdensity, 10.8$\sigma$, is higher than
in panel (a).
Panel (c): CMD for the stars at $r < 2'$, where the error bars show a typical measurement error
in color at each $r$ magnitude.
Panel (d): the same as (c) but including an isochrone (red line) for an old, metal-poor system
[age of 13~Gyr and metallicity of [M$/$H]$=-2.2$ at a distance modulus of $(m-M)_0 = 19.7$ mag].
The shaded area covers both the typical photometric error and likely intrinsic dispersion of the
CMD in star clusters.
Panel (e): the same as (c) but for field stars at $6' < r < 6'.33$,
which has the same solid angle. Note the absence of a main sequence turn-off.
}
\label{fig: space_cmd}
\end{figure*}

\section{Data and Method}

The HSC-SSP is an ongoing optical imaging survey, which consists of three layers
with different combinations of area and depth. Our search for new MW satellites
is based on its Wide layer, aiming to observe $\sim 1400$ deg$^2$ in five photometric
bands ($g$, $r$, $i$, $z$, and $y$), where the target 5$\sigma$ point-source limiting
magnitudes are
($g$, $r$, $i$, $z$, $y$) = (26.5, 26.1, 25.9, 25.1, 24.4) mag. In this paper,
we utilize the ($g$, $r$) data in the early HSC survey obtained before 2015 November,
covering $\sim 100$ deg$^2$ in 5 fields along the celestial equator.
The HSC data are processed with hscPipe v4.0.1, a branch of the Large Synoptic Survey
Telescope pipeline \citep{Ivezic2008,Juric2015} calibrated against PanSTARRS1
photometry and astrometry \citep{Schlafly2012,Tonry2012,Magnier2013}.

We use the {\it extendedness} parameter from the pipeline to select point sources.
This parameter is computed from the ratio between PSF and cmodel fluxes,
which are measured by fitting PSF models and two-component
PSF-convolved galaxy models to the source profile, respectively \citep{Abazajian2004}.
When the ratio between these fluxes is larger than 0.985,
a source is classified as a point source.
We use the parameter measured in the $i$-band, in which the seeing is typically
the best of our five filters with a median of about $0''.6$. In particular,
the $i$-band seeing for the region around our new-found satellite is
about $0''.5$.  In order to characterize the completeness and
contamination of our star/galaxy classification, we stack the COSMOS data (COSMOS
is one of our UltraDeep fields, where we have many exposures) to the depth of
the Wide survey and compare our classification against the HST/ACS data from
\citet{leauthaud2007}.  We find that the completeness, defined here
as the fraction of objects that are classified as stars by ACS,
and correctly classified as stars by HSC,
is above 90\% at $i<22.5$, and
drops to $\sim50\%$ at $i=24.5$. On the other hand, contamination, which is
defined as the fraction of HSC-classified stars which are classified as galaxies by ACS,
is close to zero at $i<23$, but increases to
$\sim 50\%$ at $i=24.5$. Based on this test, we choose to use the extendedness
parameter down to $i=24.5$ to select stars in this work\footnote{Another method
for star/galaxy classification by combining the colors of the sources
(Garmilla et al. in prep.) has also been applied
and we have confirmed that the main results of this work remain unchanged.
The full description for the analysis of the data based on this alternative
scheme will be presented in a future paper.}.
We further apply a $g - r < 1.0$ cut to eliminate numerous M-type disk stars.

In order to search for the signature of new satellites, we count stars in
$0^{\circ}.05 \times 0^{\circ}.05$ bins in right ascension and declination, with an overlap of
$0^{\circ}.025$ in each direction, where $0^{\circ}.05$ corresponds to
a typical half-light diameter ($\sim 80$~pc) of an ultra-faint dwarf at a distance of 90~kpc.
We then calculate the mean density and its dispersion over all cells for each of
the Wide layer fields to search for any spatial overdensities of stars
\citep[e.g.,][]{Koposov2008,Walsh2009}. The deviation from the mean density has close to
a Gaussian distribution.
We have found one stellar overdensity with 5.5$\sigma$ in one of
the Wide layer fields. The standard deviation is estimated
separately for each survey field (covering typically 20 to 30~deg$^2$);
each field is at different Galactic coordinates. This overdensity
is centered at $(\alpha, \delta) = (180^{\circ}.04, -0^{\circ}.68)$.
As Figure \ref{fig: space} shows, there is no corresponding overdensity in
extended objects (galaxies)\footnote{Another high-sigma overdensity (6.8$\sigma$) of the
sources with $extendedness = 0$ has been identified in the survey region, but this appears
an artefact related to scattered light from a nearby bright star.}.

In Figure \ref{fig: space_cmd}(a), we plot the spatial distribution of the stars
around this overdensity, which shows a localized concentration of stars within a circle of
radius $2'$. To get further insights into this overdensity, in Figure \ref{fig: space_cmd}(c),
we plot the $(g-r, r)$ color-magnitude diagram (CMD) of stars within the $2'$ radius circle
shown in  Figure \ref{fig: space_cmd}(a). This CMD shows signatures of main sequence (MS) stars
near its turn off (MSTO) as well as stars on the red giant branch (RGB), whereas these
features disappear when we plot stars at $6' < r < 6'.33$ with the same solid angle,
i.e. likely field stars outside the overdensity, as shown in Figure \ref{fig: space_cmd}(e).
To investigate the distribution of
the overdensity in the CMD further, we adopt a fiducial locus of stars in a typical
ultra-faint dwarf galaxy based on a PARSEC isochrone \citep{Bressan2012}, in which
we assume an age of 13~Gyr and metallicity of $z = 0.0001$ ([M$/$H]$=-2.2$).
We first derive this isochrone in the
SDSS filter system and then convert to the HSC filter system using the following
formula calibrated from both filter curves and spectral atlas of stars \citep{Gunn1983},
$g = g_{\rm SDSS} - a (g_{\rm SDSS} - r_{\rm SDSS}) - b$ and
$r = r_{\rm SDSS} - c (r_{\rm SDSS} - i_{\rm SDSS}) - d$, where
$(a,b,c,d)=(0.074,0.011,0.004,0.001)$ and the subscript SDSS denotes the SDSS system. 
This isochrone, at the assumed distance modulus of $(m-M)_0 = 19.7$ mag as determined below,
is shown in Figure \ref{fig: space_cmd}(d),
which does a good job of tracing the distributions of MSTO and RGB stars. To test the statistical
significance of the overdensity along this isochrone, we set the selection filter
defined by the CMD envelope [shaded region in Figure \ref{fig: space_cmd}(d)],
which consists of the above isochrone, 1$\sigma$ $(g-r)$ color measurement error
as a function of $r$-band magnitude, and a typical color dispersion of about $\pm 0.05$ mag
at the location of the RGB arising from a metallicity dispersion of $\pm 0.7$ dex for dSph stars.
By passing this filter over the stars in the relevant
region, we derive an overdensity that peaks at a distance modulus of 19.7~mag at a statistical
significance of 10.8$\sigma$, much higher than without the filter.
Figure \ref{fig: space_cmd}(b) shows the distribution of the stars that pass
this filter, revealing a higher overdensity contrast than Figure \ref{fig: space_cmd}(a).
This suggests that the overdensity we have found here is indeed an old stellar
system, either a globular cluster or dwarf galaxy. Hereafter we refer to this system
as Virgo~I\footnote{This is not to be confused with the so-called Virgo overdensity,
which is closer at $\sim 6$ to 20 kpc
and covering a much larger volume \citep{Juric2008}.}.
The stars selected by this isochrone filter lie along
a clear stellar sequence even in a 2-color ($g-r$, $r-i$) diagram.
We note that the statistical significance of this overdensity before
(after) passing this isochrone filter remains basically unchanged when we adopt
different magnitude limits for the sample:
5.6$\sigma$ (10.3$\sigma$) for $i < 24$~mag and
4.8$\sigma$ (9.6$\sigma$) for $i < 25$~mag.

\section{Properties of Stellar Population}

We estimate the basic structural properties of Virgo~I.
For this purpose, we adopt six parameters $(\alpha_0, \delta_0, \theta, \epsilon,
r_h, N_{\ast})$: $(\alpha_0, \delta_0)$ for the celestial coordinates
of the centroid of the overdensity, $\theta$ for its position angle from north to
east, $\epsilon$ for the ellipticity, $r_h$ for the half-light radius, and
$N_{\ast}$ for the number of stars belonging to the overdensity. 
The maximum likelihood method of \citet{Martin2008} is applied to the stars
within a circle of radius $20'$ passing the isochrone filter;
the results are summarized in Table~\ref{tab: 1}.

\begin{deluxetable}{lc}
\tablecaption{Properties of Virgo~I\label{tab:1}}
\tablewidth{0pt}
\tablehead{ \colhead{Parameter\tablenotemark{a}} & \colhead{Value}}
\startdata
Coordinates (J2000)           &
       12$^{\rm h}$00$^{\rm m}$09$^{\rm s}$.6, $-$0$^{\circ}$40$'$48$''$  \\
Galactic Coordinates ($l,b$)  & 276$^{\circ}$.94, 59$^{\circ}$.58         \\
Position angle                & $+51^{+18}_{-40}$ deg                     \\
Ellipticity                   & 0.44$^{+0.14}_{-0.17}$                    \\
$A_V$                         & 0.066 mag                                 \\
$(m-M)_0$                     & 19.7$^{+0.3}_{-0.2}$ mag                  \\
Heliocentric distance         & 87$^{+13}_{-8}$ kpc                       \\
Half light radius, $r_h$      & $1'.5 \pm 0'.4$ or 38$^{+12}_{-11}$ pc    \\
$M_{{\rm tot},V}$             & $-0.8 \pm 0.9$ mag
\enddata
\tablenotetext{a}{Integrated magnitudes are
corrected for the mean Galactic foreground extinction, $A_V$
\citep{Schlafly2011}.}
\label{tab: 1}
\end{deluxetable}

\begin{figure}[t!]
\centering
\includegraphics[width=80mm]{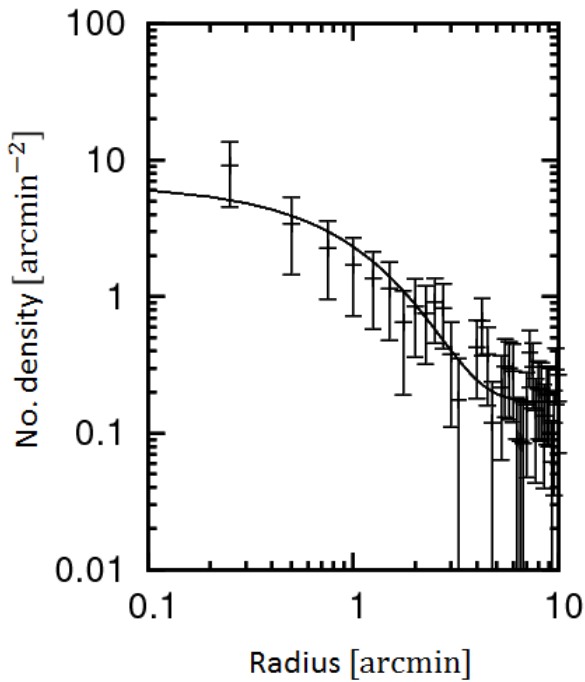}
\caption{Density profile of the stars in Virgo~I that pass the isochrone filter
shown in Figure \ref{fig: space_cmd}(b), in elliptical annuli as a function of mean radius,
where the uncertainties are derived assuming Poisson statistics.
The line shows a fitted exponential profile with $r_h = 1'.5$.}
\label{fig: profile}
\end{figure}

Figure \ref{fig: profile} shows the radial profile of the stars passing the isochrone
filter [Figure \ref{fig: space_cmd}(b)] by computing the average density within
elliptical annuli. The overplotted line corresponds to the best-fit
exponential profile with a half-light radius of $r_h = 1'.5$ or 38~pc.
This spatial size is larger than the typical size of MW globular clusters
but is consistent with the scale of dwarf satellites as examined below.

The total absolute magnitude of Virgo~I, $M_V$, is estimated by summing
the luminosities of the stars within the half-light radius, $r_h$, and then
doubling the summed luminosity \citep[e.g.,][]{Sakamoto2006}.
For the transformation from $(g, r)$ to $V$, we adopt the formula in
\citet{Jordi2006} calibrated for metal-poor Population II, which are appropriate
for stars in ultra-faint dwarf galaxies.
Assuming that the distance to this stellar system is 87~kpc or $(m-M)_0 = 19.7$~mag,
we obtain $M_V = -0.17$ mag for $r_h = 1'.5$. This value varies when we
adopt different half-light radii or different distance
moduli within their 1$\sigma$ uncertainties. We find
$M_V = +0.08$ mag if we adopt $r_h = 1'.1$ and $(m-M)_0 = 19.5$ mag
and $M_V = -1.87$ mag for $r_h = 1'.9$ and and $(m-M)_0 = 20.0$ mag.
The latter case yields a much brighter $M_V$ due to the inclusion
of a bright RGB star inside the aperture.

Shot noise due to the small number of stars in Virgo~I is a significant
additional source of uncertainty in $M_V$. We quantify this and other
sources of error using a Monte Carlo method similar to that described
in \citet{Martin2008} to determine the most likely value of $M_V$ and
its uncertainty. As summarized in Table~\ref{tab: 1} for Virgo~I,
we have derived $N_{\ast} = 19 \pm 5$ at $i < 24.5$ mag,
the distance modulus of $(m-M)_0 = 19.7_{-0.2}^{+0.3}$ mag,
and we use a stellar population model with an age of 13~Gyr and
metallicity of [M$/$H]$=-2.2$. Based on
these information, we generate $10^4$ realizations of CMDs for
three different initial mass functions (IMFs);
Salpeter, Kroupa, and Chabrier (lognormal)
\citep{Salpeter1955,Kroupa2002,Chabrier2001}. We then derive
the luminosity of the stars for each CMD at $i < 24.5$ mag, taking into
account the completeness of the observed stars with HSC. Based on
this Monte Carlo simulation, we obtain the expected values of $M_V$ as
$M_V =-0.82 \pm 0.95$, $M_V =-0.81 \pm 0.91$, and
$M_V =-0.83 \pm 0.92$, for Salpeter, Kroupa, and Chabrier IMFs, respectively.
Thus, the values of $M_V$ for these different IMF models are consistent
each other, summarized as $M_V = -0.8 \pm 0.9$ mag,
and are within the 1$\sigma$ uncertainty
of $M_V$ determined above by directly counting the observed member stars.

\begin{figure}[t!]
\centering
\includegraphics[width=90mm]{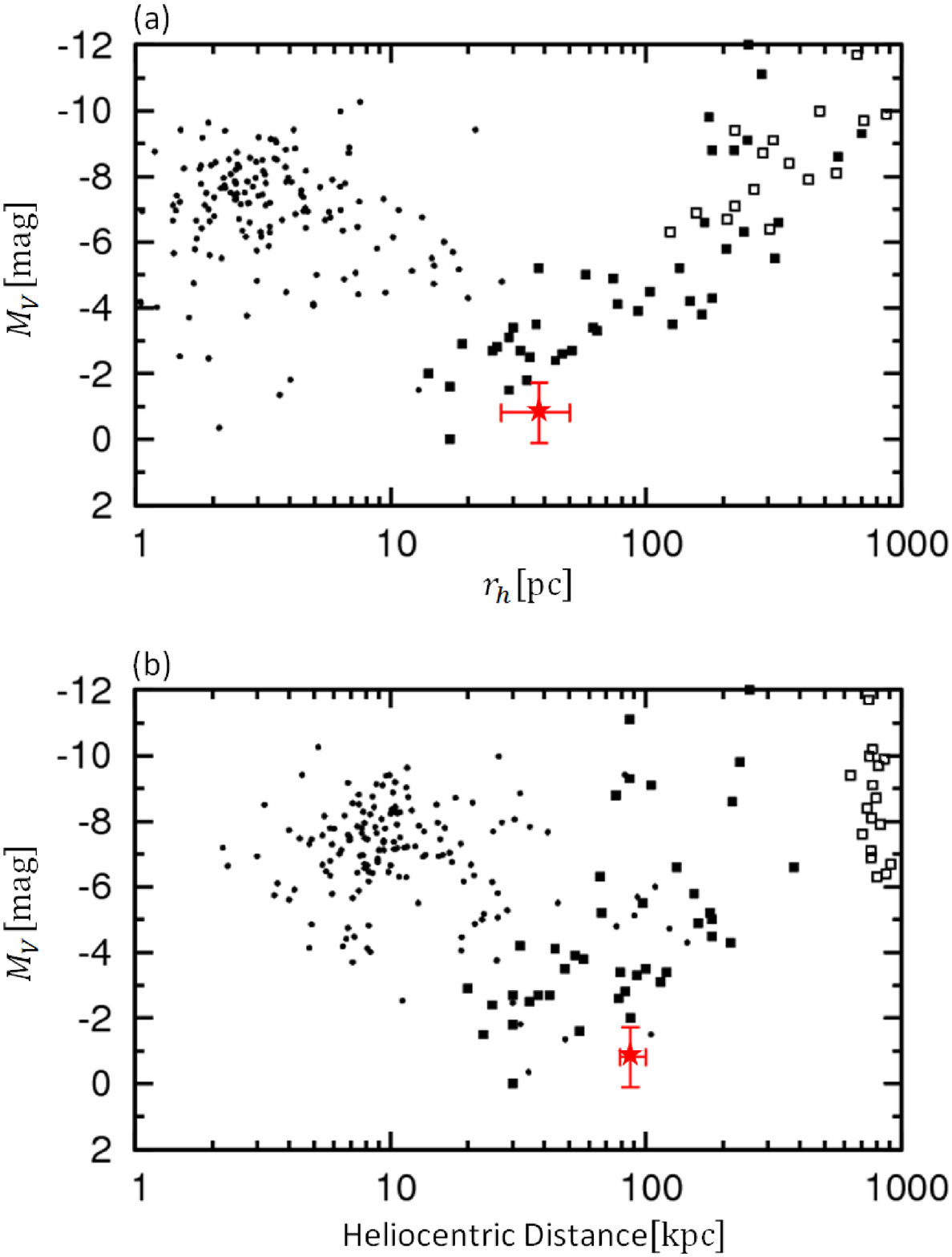}
\caption{(a) The relation between $M_V$ and $r_h$ for stellar systems. Dots denote
globular clusters in the MW taken from \citet{Harris1996}. Filled and open squares
denote the MW and M31 dSphs, respectively, taken from \citet{McConnachie2012},
the recent DES work for new ultra-faint MW dSphs
\citep{Bechtol2015,Koposov2015,Drlica-Wagner2015},
and other recent discoveries \citep{Laevens2014,Kim2015,KimJerjen2015,
Laevens2015a,Laevens2015b}.
The red star with error bars corresponds to the overdensity described in this paper,
Virgo~I, which lies within the locus defined by dSphs.
(b) The relation between $M_V$ and heliocentric distance for the systems shown
in panel (a).
}
\label{fig: abs_size}
\end{figure}

We note that the above models suggest the ratio between the number of RGB$+$HB and
that of MSTO stars is about 0.2, whereas the observed ratio is about 0.4.
This discrepancy by a factor of 2 can be understood when we consider the
contamination of some field RGBs and/or incompleteness of
faint MSTO stars.

\section{Discussion}

To assess if Virgo~I identified here is indeed a new MW dwarf satellite galaxy,
we compare its size quantified by $r_h$ with globular clusters with comparable
luminosity, in the range of $M_V \sim +0.10$ to $-1.72$ mag.
In Figure~\ref{fig: abs_size}(a), we plot the relation
between $M_V$ and $r_h$ for the MW globular clusters (dots) taken from \citet{Harris1996},
and dwarf galaxies in the MW (filled squares) and M31 (open squares) from
\citet{McConnachie2012}, the recent DES work
\citep{Bechtol2015,Koposov2015,Drlica-Wagner2015},
and other recent discoveries \citep{Laevens2014,Kim2015,KimJerjen2015,
Laevens2015a,Laevens2015b}.
The red star with error bars shows Virgo~I detected in this work.

As is clear from the figure, the current stellar system is systematically larger
than MW globular clusters with comparable $M_V$ and is located along
the locus of the MW and M31 dwarf galaxies. This is the case even if we adopt
the brighter estimate of $M_V = -1.72$ mag by considering the 1$\sigma$
uncertainty in $M_V$. Thus, 
the overdensity of the stars we have found here is a candidate ultra-faint dwarf galaxy.
This is also supported from its non-zero ellipticity of
$\epsilon = 0.44_{-0.17}^{+0.14}$, which is more similar to those of dwarf galaxies than
globular clusters.

The heliocentric distance to Virgo~I is $D = 87^{+13}_{-8}$~kpc,
where the error estimate is derived from the range of the distance
yielding the 1$\sigma$ decrease in the statistical significance of Virgo~I after
passing the isochrone filter [defined in Figure \ref{fig: space_cmd}(d)]
from its peak value of 10.8$\sigma$.
This distance is beyond
the reach of previous surveys for MW dwarfs with comparable luminosity.
This is demonstrated in Figure~\ref{fig: abs_size}, which shows the relation between $M_V$ and $D$
for the MW and M31 dwarfs as well as the MW globular clusters.

\section{Conclusions}

We have identified a new ultra-faint dwarf satellite of the MW, Virgo~I,
in the constellation of Virgo. The satellite is located at a heliocentric
distance of 87~kpc and its absolute magnitude in the $V$ band is estimated 
as $M_V = -0.8 \pm 0.9$ mag, which is comparable to or fainter than
that of the faintest dwarf satellite, Segue~1. The half-light radius of Virgo~I
is estimated to be $\sim 38$ pc, significantly larger than
globular clusters with the same luminosity, 
suggesting that it is a dwarf galaxy. To set further constraints on Virgo~I,
follow-up spectroscopic studies of bright RGB stars will be useful to investigate
their membership and to determine the chemical and dynamical properties
in this dwarf satellite.

Virgo~I is located beyond the reach of the SDSS: its limiting magnitude
of $r = 22.2$ implies that the completeness radius beyond which a faint dwarf galaxy
like Virgo~I will not be detected \citep{Tollerud2008} is 28~kpc.
With Subaru/HSC, this completeness radius for Virgo~I is estimated as 89~kpc,
if we adopt the limiting $i$-band magnitude of 24.5 mag combined with a typical
$(r-i)$ color of $\simeq 0.2$. Thus, Virgo~I with $D = 87^{+13}_{-8}$ kpc is 
located just at the edge where Subaru/HSC can reach.
We therefore expect the presence of yet unidentified faint satellites in the outer parts of
the MW halo as the HSC survey continues. Deep imaging surveys for these faint and distant
satellites are indeed important to get further insights into their true number
and thus the nature of dark matter on small scales.

\acknowledgments
We thank the referee for his/her helpful comments and suggestions.
This work is supported in part by JSPS Grant-in-Aid for Scientific 
Research (B) (No. 25287062) and MEXT Grant-in-Aid for Scientific Research on 
Innovative Areas (No. 15H05889).

The Hyper Suprime-Cam (HSC) collaboration includes the astronomical
communities of Japan and Taiwan, and Princeton University.  The HSC
instrumentation and software were developed by the National
Astronomical Observatory of Japan (NAOJ), the Kavli Institute for the
Physics and Mathematics of the Universe (Kavli IPMU), the University
of Tokyo, the High Energy Accelerator Research Organization (KEK), the
Academia Sinica Institute for Astronomy and Astrophysics in Taiwan
(ASIAA), and Princeton University.  Funding was contributed by the FIRST 
program from Japanese Cabinet Office, the Ministry of Education, Culture, 
Sports, Science and Technology (MEXT), the Japan Society for the 
Promotion of Science (JSPS),  Japan Science and Technology Agency 
(JST),  the Toray Science  Foundation, NAOJ, Kavli IPMU, KEK, ASIAA,  
and Princeton University.
This paper makes use of software developed for the Large Synoptic Survey Telescope. We thank the
LSST Project for making their code freely available. The Pan-STARRS1 (PS1) Surveys have been made
possible through contributions of the Institute for Astronomy, the University of Hawaii, the Pan-STARRS
Project Office, the Max-Planck Society and its participating institutes, the Max Planck Institute for
Astronomy and the Max Planck Institute for Extraterrestrial Physics, The Johns Hopkins University,
Durham University, the University of Edinburgh, Queen's University Belfast, the Harvard-Smithsonian
Center for Astrophysics, the Las Cumbres Observatory Global Telescope Network Incorporated, the
National Central University of Taiwan, the Space Telescope Science Institute, the National Aeronautics
and Space Administration under Grant No. NNX08AR22G issued through the Planetary Science Division
of the NASA Science Mission Directorate, the National Science Foundation under Grant
No.AST-1238877, the University of Maryland, and Eotvos Lorand University (ELTE).


\end{document}